\documentclass[aps,pra,twocolumn,floatfix]{revtex4}
\usepackage{graphicx}
\usepackage{nicefrac}
\usepackage{amsmath}
\usepackage{amsfonts}
\usepackage{amssymb}
\usepackage{amsthm}
\usepackage{epsf}
\usepackage{bm}
\usepackage{bbm}
\usepackage{longtable}

\usepackage{dcolumn}

\newcommand{\cE}{{\cal E}}

\usepackage{color}

\newcolumntype{.}{D{x}{}{-1}}

\begin{document}

\newcommand{\half}{\frac12}
\newcommand{\vare}{\varepsilon}
\newcommand{\pr}{^{\prime}}
\newcommand{\ppr}{^{\prime\prime}}
\newcommand{\pp}{{p^{\prime}}}
\newcommand{\hp}{\hat{\bfp}}
\newcommand{\hpp}{\hat{\bfpp}}
\newcommand{\hx}{\hat{\bfx}}
\newcommand{\hq}{\hat{\bfq}}
\newcommand{\hz}{\hat{\bfz}}
\newcommand{\hr}{\hat{\bfr}}
\newcommand{\hn}{\hat{\bfn}}
\newcommand{\rx}{{\rm x}}
\newcommand{\rp}{{\rm p}}
\newcommand{\rmq}{{\rm q}}
\newcommand{\rpp}{{{\rm p}^{\prime}}}
\newcommand{\rk}{{\rm k}}
\newcommand{\bfe}{{\bm e}}
\newcommand{\bfp}{{\bm p}}
\newcommand{\bfpp}{{\bm p}^{\prime}}
\newcommand{\bfq}{{\bm q}}
\newcommand{\bfx}{{\bm x}}
\newcommand{\bfk}{{\bm k}}
\newcommand{\bfz}{{\bm z}}
\newcommand{\bfr}{{\bm r}}
\newcommand{\bfn}{{\bm n}}
\newcommand{\bphi}{{\mbox{\boldmath$\phi$}}}
\newcommand{\bnabla}{\mbox{\boldmath$\nabla$}}
\newcommand{\bgamma}{{\mbox{\boldmath$\gamma$}}}
\newcommand{\balpha}{{\mbox{\boldmath$\alpha$}}}
\newcommand{\bsigma}{{\mbox{\boldmath$\sigma$}}}
\newcommand{\bomega}{{\mbox{\boldmath$\omega$}}}
\newcommand{\bvare}{{\mbox{\boldmath$\varepsilon$}}}

\newcommand{\bGamma}{\bm{\Gamma}}
\newcommand{\intzo}{\int_0^1}
\newcommand{\intinf}{\int^{\infty}_{-\infty}}
\newcommand{\ka}{\kappa_a}
\newcommand{\kb}{\kappa_b}
\newcommand{\ThreeJ}[6]{
        \left(
        \begin{array}{ccc}
        #1  & #2  & #3 \\
        #4  & #5  & #6 \\
        \end{array}
        \right)
        }
\newcommand{\SixJ}[6]{
        \left\{
        \begin{array}{ccc}
        #1  & #2  & #3 \\
        #4  & #5  & #6 \\
        \end{array}
        \right\}
        }
\newcommand{\NineJ}[9]{
        \left\{
        \begin{array}{ccc}
        #1  & #2  & #3 \\
        #4  & #5  & #6 \\
        #7  & #8  & #9 \\
        \end{array}
        \right\}
        }
\newcommand{\Dmatrix}[4]{
        \left(
        \begin{array}{cc}
        #1  & #2   \\
        #3  & #4   \\
        \end{array}
        \right)
        }
\newcommand{\cross}[1]{#1\!\!\!/}
\newcommand{\eps}{\epsilon}
\newcommand{\beq}{\begin{equation}}
\newcommand{\eeq}{\end{equation}}
\newcommand{\beqn}{\begin{eqnarray}}
\newcommand{\eeqn}{\end{eqnarray}}
\newcommand{\lbr}{\langle}
\newcommand{\rbr}{\rangle}
\newcommand{\Za}{Z\alpha}

\title{Theory of the helium isotope shift}

\author{Krzysztof Pachucki} \affiliation{Institute of Theoretical
        Physics, University of Warsaw, Pasteura 5, 02--093 Warsaw, Poland}

\author{V.~A. Yerokhin}
\affiliation{Center for Advanced Studies, St.~Petersburg State Polytechnical University,
Polytekhnicheskaya 29,
        St.~Petersburg 195251, Russia}

\begin{abstract}
Theory of the isotope shift of the centroid energies of light few-electron atoms is reviewed.
Numerical results are presented for the isotope shift of the $2^3P$-$2^3S$ and $2^1S$-$2^3S$
transition energies of $^3$He and $^4$He. By comparing theoretical predictions for the isotope
shift with the experimental results, the difference of the squares of the nuclear charge radii of
$^3$He and $^4$He, $\delta R^2$, is determined with high accuracy.
\end{abstract}

\maketitle

\section{Introduction}

Helium is the simplest few-electron atom that can be calculated {\em ab initio}
\cite{morton:06:cjp} and measured with a very high accuracy. Optical measurements in $^3$He
\cite{rooij:11,cancio:12:3he} and $^4$He \cite{cancio:04} has recently reached the relative
precision of few parts in $10^{-12}$. Such measurements are sensitive to the nuclear size effect on
the level of $10^{-4}$. They are approaching the level that might allow an improved determination
of the Rydberg constant, which is currently considered as the one of the best known fundamental
constants \cite{mohr:12:codata}. The accuracy of the present-day experiments is high enough for an
accurate spectroscopic determination of the nuclear charge radii of $^3$He and $^4$He. Such
determination would be of particular interest in view of the discrepancy for the proton charge
radius observed in the muonic hydrogen experiment \cite{pohl:10,antognini:13}. The follow-up
experiment on the muonic helium \cite{antognini:11} might soon bring the charge radius of helium
into focus.

Unfortunately, the present-day theory of the energy levels in helium
\cite{pachucki:06:he,yerokhin:10:helike} is not good enough to provide the nuclear charge radii of
$^3$He and $^4$He with the required precision. A more feasible task at present is the
high-precision determination of the nuclear charge radii difference of these isotopes, which is
extracted from the isotope shift of the transition energies. In the current paper we make a review
of the theory of the isotope shift in light few-electron atoms, describe our calculation of the
isotope shift for helium, and determine the difference of the squares of the nuclear charge radii
of $^3$He and $^4$He, $\delta R^2$. The results of the $\delta R^2$ determination have already been
published in Ref.~\cite{cancio:12:3he}. In this work we present details of the calculation and the
underlying theory.

We define the {\em isotope shift} of two isotopes as the difference of the centroid energies. More
specifically, we are presently interested in the centroid energies of the $2P$ and $2S$ states,
which are defined as an average over all fine and hyperfine energy sublevels,
\begin{eqnarray} \label{centroid}
  E(2^{2S+1}L) &=&\frac{\sum_{J, F} (2\,F+1)\,E(2^{2S+1}L_{J,F})}
                {(2\,I+1)\,(2\,S+1)\,(2\,L+1)}\,,
\end{eqnarray}
where $^{2S+1}L$ denotes the state with the angular momentum $L$ and the spin $S$. This definition
of the isotope shift differs from the definition sometimes used previously (e.g., in
Refs.~\cite{shiner:95,morton:06:cjp,rooij:11}) by the fact that we average out not only the
hyperfine but also the fine-structure splitting. The advantage of using the centroid energy is that
its theory is much more transparent than that of the individual sublevels $2^{2S+1}L_{J,F}$.
Moreover, the centroid transition energy can be directly accessed in an experiment, by measuring
all independent fine and hyperfine transitions and averaging them out.

Theory of the isotope shift is much simpler that theory of the energy levels because only very
restricted set of operators contribute to it. Specifically, only the operators that depend on the
nuclear mass or the nuclear size contribute to the isotope shift. In addition, due to the averaging
over the fine and hyperfine energy sublevels in (\ref{centroid}), all operators that depend the
nuclear or the electron spin do not (almost) contribute to the isotope shift. More exactly, such
operators contribute only through the second-order perturbations, most of which are negligible at
the level of the present  interest.

\section{Theory}

Within the nonrelativistic QED approach, energy levels of atoms are represented by an expansion in
powers of the fine-structure constant $\alpha$,
\begin{equation} \label{e1} E(\alpha) = E^{(2)} + E^{(4)} + E^{(5)} + E^{(6)}
  + E^{(7)} + \ldots,
\end{equation}
where $E^{(n)}\equiv m\alpha^n\cE^{(n)}$ is a contribution of order $\alpha^n$ and may include
powers of $\ln\alpha$. Each of $\cE^{(n)}$ is in turn expanded in powers of the electron-to-nucleus
mass ratio $m/M$
\begin{equation} \label{e2} \cE^{(n)} = \cE^{(n)}_{\infty}+ \cE^{(n)}_M +
  \cE^{(n)}_{M^2}+\ldots\,,
\end{equation}
where $\cE^{(n)}_M$ denotes the correction of first order in $m/M$ and $\cE^{(n)}_{M^2}$ is the
second-order correction. For the description of the isotope shift, we need only the part that
depends on the nuclear mass $M$, i.e., corrections $\cE^{(n)}_M$, $\cE^{(n)}_{M^2}$, etc.

\subsection{Nonrelativistic recoil}

The leading term in the expansion (\ref{e1}) is the nonrelativistic energy $E^{(2)}$, which is the
eigenvalue of the nonrelativistic Hamiltonian,
\begin{equation}
H^{(2)}_M = \frac{\vec P^{\,2}}{2\,M} +
\sum_a \frac{\vec p_a^{\,\,2}}{2\,m} -
\sum_a
  \frac{Z}{r_a} + \sum_{a<b} \frac1{r_{ab}}\,
\end{equation}
where $Z$ and $\vec P$ are the nuclear charge and momentum, respectively; the indexes $a$ and $b$
numerate the electrons, $\vec p_a$ is the momentum of the electron $a$, $r_a$ is the distance
between the electron $a$ and the nucleus and $r_{ab}$ is the radial distance between the electrons
$a$ and $b$.

Assuming that the total momentum is zero,
\begin{equation}
\vec P^2 = \biggl( - \sum_a \vec p_a \biggr)^2 = \sum_a \vec p_a^{\,\,2} + 2 \sum_{a<b} \vec p_a \cdot \vec p_b\,,
\end{equation}
and rescaling momenta and position operators in units of the reduced mass $\mu = mM/(m+M)$, $ \vec
p_a \rightarrow \mu\,\vec p_a$ and $\vec r_a \rightarrow \vec r_a/\mu$, the Hamiltonian $H^{(2)}_M$
is represented as a sum of two terms,
\begin{equation} \label{eq0:a}
H^{(2)}_M = \mu\,H^{(2)}_{\infty} + \frac{\mu^2}{m}\,H^{(2)}_{{\rm rec}}\,,
\end{equation}
with
\begin{equation}
H^{(2)}_{\infty} = \sum_a \biggl(\frac{\vec p_a^{\,\,2}}{2} -
  \frac{Z}{r_a}\biggr) + \sum_{a<b} \frac1{r_{ab}}
\end{equation}
and
\begin{equation} \label{MP} H^{(2)}_{{\rm rec}} = \frac{m}{M}\,
  \sum_{a<b} \vec{p}_a\cdot\vec{p}_b\,.
\end{equation}
The eigenvalue of $H^{(2)}_{\infty}$ is ${\cal E}^{(2)}_\infty$, the nonrelativistic energy in the
nonrecoil limit. In the present work we are interested in the recoil (i.e., $M$-dependent)
corrections to $\cE^{(2)}_{\infty}$. The correction coming from $\mu\,H^{(2)}_{\infty}$ is often
termed as the normal mass shift. It is given by
\begin{align} \label{nr0}
  \cE^{(2)}_{M,0} = \left(\frac{\mu}{m}-1\right)\, \cE^{(2)}_{\infty}\,.
\end{align}
The second recoil correction is induced by $H^{(2)}_{{\rm rec}}$; it is known as the specific mass
shift or the mass polarization. This correction can, in principle, be calculated numerically to all
orders in $m/M$ by including $H^{(2)}_{{\rm rec}}$ into the nonrelativistic Hamiltonian,
determining the corresponding eigenvalue and subtracting the nonrecoil limit $M\to \infty$. This
approach works very well for $P$ states but becomes numerically unstable for $S$ states (with the
basis of the fully correlated exponential functions used in the present work). Because of this, we
obtain the specific mass shift corrections by perturbation theory, extending it up to the third
order. The perturbation corrections are
\begin{align} \label{nr1}
  \cE^{(2)}_{M,1} = \left(\frac{\mu}{m}\right)^2\,\left< H^{(2)}_{{\rm rec}}
  \right>\,,
\end{align}
\begin{align} \label{nr2}
  \cE^{(2)}_{M,2} = \left(\frac{\mu}{m}\right)^3\,
\left< H^{(2)}_{{\rm rec}}\, \frac1{(\cE_0-H_0)'}\, H^{(2)}_{{\rm rec}}\right>\,,
\end{align}
\begin{align} \label{nr3}
  \cE^{(2)}_{M,3}  = \left(\frac{\mu}{m}\right)^4\,\biggl[ &\
\left< H^{(2)}_{{\rm rec}}\, \frac1{(\cE_0-H_0)'}\, H^{(2)}_{{\rm rec}} \, \frac1{(\cE_0-H_0)'}\, H^{(2)}_{{\rm rec}}\right>
  \nonumber \\ &
- \left< H^{(2)}_{{\rm rec}}\right>
\left< H^{(2)}_{{\rm rec}}\, \frac1{{(\cE_0-H_0)'}^2}\, H^{(2)}_{{\rm rec}}\right>
\biggr]\,,
\end{align}
where $H_0 \equiv H^{(2)}_{\infty}$ is the nonrelativistic Hamiltonian in the nonrecoil limit and
$\cE_0\equiv \cE^{(2)}_{\infty}$ is its eigenvalue. We checked that, for $P$ states, the
perturbation expansion results are in a very good agreement with the nonpertubative treatment.

\subsection{Relativistic recoil}

The leading relativistic correction to the energy is given by the expectation value of the
Breit-Pauli Hamiltonian $H^{(4)}$ \cite{bethesalpeter}. The Breit-Pauli Hamiltonian with the full
reduced-mass dependence is given by:
\begin{align} \label{H4}
H^{(4)} &\ = \left(\frac{\mu}{m}\right)^3 \sum_a \biggl[-\frac{\mu}{m}\frac{\vec p^{\,4}_a}{8} +
  \frac{ \pi Z}{2}\,\delta^3(r_a) +\frac{Z}{4}\, \vec\sigma_a\cdot\frac{\vec
    r_a}{r_a^3}\times \vec p_a\biggr]
  \nonumber \\
  & +\left(\frac{\mu}{m}\right)^3 \sum_{a<b}\biggl\{ -\pi\, \delta^3(r_{ab}) -\frac1{2}\, p_a^i\,
  \biggl(\frac{\delta^{ij}}{r_{ab}}+\frac{r^i_{ab}\,r^j_{ab}}{r^3_{ab}}
  \biggr)\, p_b^j \nonumber \\ & - \frac{2 \pi}{3}\,\vec\sigma_a
  \cdot\vec\sigma_b\,\delta^3(r_{ab}) +\frac{\sigma_a^i\,\sigma_b^j} {4
    r_{ab}^3}\,
  \biggl(\delta^{ij}-3\,\frac{r_{ab}^i\,r_{ab}^j}{r_{ab}^2}\biggr) \nonumber
  \\ & +\frac1{4\,r_{ab}^3} \bigl[ 2\,\bigl(\vec\sigma_a\cdot\vec
  r_{ab}\times\vec p_b - \vec\sigma_b\cdot\vec r_{ab}\times\vec p_a\bigr)
  \nonumber \\ & + \bigl(\vec\sigma_b\cdot\vec r_{ab}\times\vec p_b -
  \vec\sigma_a\cdot\vec r_{ab}\times\vec p_a\bigr)\bigr]\biggr\}\,,
\end{align}
where $\vec \sigma_a$ is the vector of Pauli $\sigma$ matrices acting on the $a$ electron. The
Breit Hamiltonian without the reduced mass dependence (i.e., with $\mu \to m$) will be denoted by
$H^{(4)}_{\infty}$. %, and its expectation value, by $\cE^{(4)}_{\infty}$.

We are presently interested in the recoil corrections to the Breit contribution. First of all,
there is a recoil correction induced by the reduced mass in Eq.~(\ref{H4}),
\begin{align} \label{br0}
\cE^{(4)}_{M,0} = \left< H^{(4)} - H^{(4)}_{\infty}\right>\,.
\end{align}

The first-order recoil correction $\cE^{(4)}_{M,1}$ is induced by the recoil addition to the Breit
Hamiltonian and the second-order perturbation correction of the non-recoil Breit Hamiltonian
$H^{(4)}_{\infty}$ and the mass-polarization operator $H^{(2)}_{{\rm rec}}$,
\begin{align}\label{br1}
\cE^{(4)}_{M,1} = \left(\frac{\mu}{m} \right)^3\,\left< H^{(4)}_{\rm rec} \right> + 2 \left(\frac{\mu}{m} \right)\,\left<H^{(4)}_{\infty}\,\frac1{{(\cE_0-H_0)'}}\, H^{(2)}_{{\rm rec}}\right>\,,
\end{align}
where the recoil addition to the Breit Hamiltonian is
\begin{align} \label{fsrec} H^{(4)}_{\rm rec} = &\ \frac{Zm}{2M}\, \sum_{ab}
  \biggl[\frac{\vec r_a}{r_a^3}\times \vec p_b \cdot\vec \sigma_a -
  p_a^i\,\left(\frac{\delta^{ij}}{r_a}+\frac{r^i_ar^j_a}{r_a^3}\right) p_b^j
  \biggr] \,.
\end{align}
The first term in the above formula represents the electron-nucleus spin-orbit interaction, whereas
the second term is induced by the electron-nucleus orbit-orbit interaction.

The second-order recoil correction $\cE^{(4)}_{M,2}$ consists of several parts,
\begin{align}\label{br2}
\cE^{(4)}_{M,2} = \cE^{(4)}_{M,2a} + \cE^{(4)}_{M,2b} + \cE^{(4)}_{M,2c}\,.
\end{align}
The first one is the third-order perturbation correction of the non-recoil Breit Hamiltonian
$H^{(4)}_{\infty}$ and two mass-polarization operators $H^{(2)}_{{\rm rec}}$,
\begin{align}
\cE^{(4)}_{M,2a} &\ = 2 \left<H^{(4)}_{\infty}\,\frac1{{(\cE_0-H_0)'}}\, H^{(2)}_{{\rm rec}}\,\frac1{{(\cE_0-H_0)'}}\, H^{(2)}_{{\rm rec}}\right>
 \nonumber \\ &
+ \left< H^{(2)}_{{\rm rec}}\,\frac1{{(\cE_0-H_0)'}}\, H^{(4)}_{\infty}\,\frac1{{(\cE_0-H_0)'}}\, H^{(2)}_{{\rm rec}}\right>
 \nonumber \\ &
- \left< H^{(4)}_{\infty} \right> \left< H^{(2)}_{{\rm rec}}\,\frac1{{(\cE_0-H_0)'}^2}\, H^{(2)}_{{\rm rec}}\right>
 \nonumber \\ &
- 2\left< H^{(2)}_{\rm rec} \right> \left< H^{(4)}_{\infty}\,\frac1{{(\cE_0-H_0)'}^2}\, H^{(2)}_{{\rm rec}}\right>\,.
\end{align}
The second part is the second-order perturbation correction of the recoil Breit Hamiltonian
$H^{(4)}_{\rm rec}$ and the mass-polarization operators $H^{(2)}_{{\rm rec}}$,
\begin{align}
\cE^{(4)}_{M,2b} = 2 \left<H^{(4)}_{\rm rec}\,\frac1{{(\cE_0-H_0)'}}\, H^{(2)}_{{\rm rec}}\right>\,.
\end{align}
The third part is the Darwin term, which reads for the $I = 1/2$ nuclei,
\begin{align} \label{br2end}
\cE^{(4)}_{M,2c} = \left( \frac{m}{M}\right)^2\, \left< \frac{Z\pi}{2}\, \sum_a \delta(\vec r_a) \right>\,.
\end{align}
The Darwin term is zero for spinless nuclei.

\subsection{QED recoil}

The leading QED correction $\cE^{(5)}_{\infty}$ in the nonrecoil limit is given by
\cite{araki:57,sucher:58}
\begin{align} \label{E5} \cE^{(5)}_{\infty} &\ =
  \frac{4Z}{3} \,\biggl[ \ln (\Za)^{-2}\,
   +  \frac{19}{30}-\ln \left( \frac{k_0}{Z^2}\right)\biggr]\,
  \sum_a \langle \delta^3(r_a) \rangle
 \nonumber \\ &
 + \biggl[ \frac{14}{3}\ln
  (\Za)\, + \frac{164}{15} \biggr]\, \sum_{a<b} \langle \delta^3(r_{ab}) \rangle
 \nonumber \\ &
  -\frac{14}{3}\,\sum_{a<b}\left< \frac1{4\pi
      r_{ab}^3}+\delta^3(r_{ab})\,\ln Z\right>
\,,
\end{align}
where the singular operator $r^{-3}$ is defined by
\begin{align} \label{rcube} \left\langle\frac{1}{r^3}\right\rangle &\ \equiv
  \lim_{a\rightarrow 0}\int d^3 r\, \phi^{*}(\vec r)\,\phi(\vec r) \nonumber
  \\ & \times \left[\frac{1}{r^3}\,\Theta(r-a) + 4\,\pi\,\delta^3(r)\,
    (\gamma+\ln a)\right]\,,
\end{align}
where $\gamma$ is the Euler constant.  The Bethe logarithm is defined as
\begin{align} \label{bethe} \ln (k_0) = \frac{\Bigl\langle\sum_a \vec
    p_a\,(H_0-\cE_0)\, \ln\bigl[2\,(H_0-\cE_0)\bigr]\, \sum_b\vec
    p_b\Bigr\rangle}{2\,\pi\,Z\, \Bigl\langle\sum_c\delta^3(r_c)\Bigr\rangle}.
\end{align}

The recoil correction $\cE^{(5)}_M$ consists of three parts \cite{pachucki:00:herec},
\begin{equation} \label{E51} \cE^{(5)}_M = \frac{m}{M}\bigl(
  \cE_{1}+\cE_{2}+\cE_{3}\bigr) \,,
\end{equation}
where
\begin{align} \label{E511} \cE_{1} = -3\, E^{(5)}_{\infty}+ \frac{4 Z}{3}
  \sum_a \langle \delta^3(r_a)\rangle - \frac{14}{3} \sum_{a<b} \langle
  \delta^3(r_{ab})\rangle \,,
\end{align}
\begin{align} \label{E512}
  \cE_{2} = &\ Z^2 \Biggl[ -\frac{2}{3}\ln (\Za)+\frac{62}{9} -\frac{8}{3}\ln
  \left(\frac{k_0}{Z^2}\right) \Biggr]\, \sum_a \langle \delta^3(r_a)\rangle
  \nonumber \\ & -\frac{14\, Z^2}{3}\,\sum_a \left< \frac1{4\pi
      r_{a}^3}+\delta^3(r_{a})\,\ln Z\right> \,,
\end{align}
and $(m/M)\,\cE_{3}$ is the first-order perturbation of $E^{(5)}_{\infty}$ due to the
mass-polarization operator $H^{(2)}_{\rm rec}$.

\subsection{Higher-order recoil corrections}

The QED correction of order $m^2\alpha^6 /M$ was estimated by using the hydrogen results
\cite{mohr:12:codata},
\begin{align}
\cE^{(6)}_{M, \rm QED} &\ = Z^2\pi \, \label{E61}
    \sum_a \langle \delta^3(r_a)\rangle
\biggl[ \left(\frac{\mu^3}{m^3}-1\right) \left( \frac{427}{96}-2\ln 2\right)
 \nonumber \\ &
    + \frac{m}{M}\,\left(\frac{35}{36} - \frac{448}{27 \pi^2} -2 \ln 2 + \frac{6}{\pi^2}\,\zeta(3)\right)
 \nonumber \\ &
    + Z\, \frac{m}{M}\,\left(4 \ln 2 -\frac{7}{2}\right) \biggr] \,.
\end{align}
The first term in the brackets in the above equation is due to the reduced mass scaling of the
$A_{50}$ one-loop QED contribution, the second term is the radiative recoil correction, and the
third term is the pure recoil contribution.

Another important correction of order $m^2\alpha^6 /M$ comes from the second-order hyperfine
correction. We keep only the important part of this correction for the $2P$ states, which is
enhanced by the small $2^3P-2^1P$ energy difference in the denominator. This correction is referred
to as the hyperfine mixing contribution and is given by, for the $2^3P$ state,
\begin{align} \label{hfsmix}
E^{(6)}_{M, \rm hfsmix} &\ = \frac{\left|\left< 2^3P|H^{(4)}_{\rm hfs} | 2^1P\right>\right|^2}{E_0(2^3P)-E_0(2^1P)}\,,
\end{align}
where $H^{(4)}_{\rm hfs}$ is the operator responsible for the leading-order hyperfine structure
splitting. In relativistic units,
\begin{align} \label{hfs}
H^{(4)}_{\rm hfs} = \mu_B\, g\,\mu_N\,\vec{I} \cdot &\ \sum_a \biggl[
\frac{8\pi}{3}\,\delta(\vec{r}_a)\,\vec{\sigma}_a
 \nonumber \\ &
+ \frac{3 \vec{r}_a \left(\vec{\sigma}_a\cdot\vec{r}_a \right)-\vec{\sigma}_a\,r_a^2}{r_a^5}
+ 2 \frac{\vec{r}_a \times \vec{p}_a}{r_a^3}\biggr]\,,
\end{align}
where $\mu_B = e/(2m)$ is the Bohr magneton, $\mu_N = e/(2m_p)$ is the nuclear magneton (with $m_p$
being the proton mass), $g$ is the nuclear $g$ factor, and $\vec{I}$ is the nuclear spin. We note
that the last term in Eq.~(\ref{hfs}) does not contribute to the hyperfine mixing. For the $2^1P$
state, the hyperfine mixing contribution is given by the same expression (\ref{hfsmix}) with the
opposite overall sign.

\subsection{Nuclear polarizability}

The nuclear polarizability correction to the energy levels can be represented as
\begin{align}\label{np}
\cE^{(4)}_{\rm pol} = -
    \sum_a \langle \delta^3(r_a)\rangle\, \widetilde{\alpha}_{\rm pol}\,.
\end{align}
The coefficient $\widetilde{\alpha}_{\rm pol}$ in the above formula was evaluated
\cite{pachucki:07:heliumnp} as $\widetilde{\alpha}_{\rm pol}(\mbox{\rm $^4$He}) =
2.07\,(20)$~fm$^3$
 and $\widetilde{\alpha}_{\rm pol}(\mbox{\rm $^3$He}) = 3.56\,(36)$~fm$^3$. Somewhat smaller values
were later obtained in Ref.~\cite{stetcu:09}, the difference being not important at the level of
interest of the present investigation.

\subsection{Finite nuclear size}

The leading finite nuclear size (fs) correction to the centroid energy is, in relativistic units,
\begin{align} \label{fns}
E_{\rm fs} = & \ \frac{2\pi}{3}\,m\,Z\alpha\, \left(\frac{\mu}{m}\right)^3
\,\frac{R^2}{\lambdabar_C^2}\,
  \Bigl\langle \sum_a \delta^3(r_a)\Bigr\rangle
 \nonumber \\ & \times
  \biggl[
     1-(Z\alpha)^2\, \ln (Z\alpha R/\lambdabar_C) + (Z\alpha)^2 f_{\rm fs}\biggr]\,,
\end{align}
where $R$ is the root-mean-square (rms) nuclear charge radius $R = \sqrt{\langle R^2 \rangle}$,
$\lambdabar_C$ is the Compton wavelength divided by $ 2\pi$ and $f_{\rm fs}$ is the nonlogarithmic
relativistic and higher-order correction. The second term in the brackets in Eq.~(\ref{fns}) is the
leading logarithmic relativistic correction, known from the hydrogen theory \cite{mohr:12:codata}.
It is state-independent and can be directly generalized to the helium case.

The nonlogarithmic relativistic correction $f_{\rm fs}$ is state dependent and, moreover, dependent
on the nuclear model. In the present work, we estimate $f_{\rm fs}$ basing on the theory of He$^+$,
where it can be obtained by a direct numerical evaluation \cite{yerokhin:11:fns}. Our calculations,
performed with quadruple-precision arithmetics for the $1s$ state of He$^+$, yield $f_{\rm fs} =
-0.04$ for the Gauss nuclear model, $f_{\rm fs} = 0.02$ for the sphere nuclear model, and $f_{\rm
fs} = -0.14$ for the exponential-distribution nuclear model. Basing on these results, we estimate
the corresponding correction for helium as $f_{\rm fs} = 0.0(2)$. The error ascribed to $f_{\rm
fs}$ corresponds to the fractional error of $4\times 10^{-5}$ in $E_{\rm fs}$, which is negligible
at the level of the present experimental interest.

The correction beyond the mean-square charge radius $E_Z$ comes from the two-photon exchange
\cite{friar}, which is, in relativistic units,
\begin{equation}
E_Z = -\frac{\pi}{3}\,m\,(Z\,\alpha)^2\,\Bigl(\frac{\mu}{m}\Bigr)^3
\, \frac{\langle R^3\rangle}{\lambdabar_C^3}
\Bigl\langle\sum_a\delta^3(r_a)\Bigr\rangle\,.
\end{equation}
Since it is proportional to $\langle R^3\rangle$ it depends on the charge distribution. However,
this effect is almost negligible for light atoms, namely the ratio
\begin{equation}
\frac{E_Z}{E_{\rm fs}} = -\frac{Z\,\alpha}{2}\,
\frac{\langle R^3\rangle}{\lambdabar_C\,\langle R^2 \rangle}
\end{equation}
for He amounts to about $4\times 10^{-5}$.

\section{Theoretical results}

Our main goal in the present investigation is the determination of the difference of the nuclear
charge radii from the isotope shift. To this end, we separate our theory of the isotope shift into
two parts. The first part is the theory the isotope shift for the point-like nucleus. The second
part is the finite nuclear size effect, which is parameterized as $\delta E_{\rm fs} = C\,R^2$,
where $R$ is the rms nuclear charge radius and the coefficient $C$ is calculated numerically. Now,
a comparison of the experimental value with the theoretical results for the isotope shift allows us
to extract the difference of the squared rms nuclear charge radii of the two isotopes.

Numerical results of our calculation of the isotope shift of the 2$^3P$-2$^3S$ and 2$^1S$-2$^3S$
transitions for the point nucleus are presented in Tables~\ref{tab:is} and \ref{tab:is:2},
respectively. The following masses of the helion $m_{h}$ and the alpha particle $m_{\alpha}$ were
used in the calculations \cite{mohr:12:codata}, $m/m_{h}=1.819\,543\,0761\,(17) \times 10^{-4}$ and
$m/m_{\alpha}=1.370\,933\,555\,78\,(55) \times 10^{-4}$. The first four lines of the tables display
the numerical results for the nonrelativistic ($\sim \alpha^2$) recoil corrections given by
Eqs.~(\ref{nr0}), (\ref{nr1}), (\ref{nr2}), and (\ref{nr3}), respectively. The next three lines of
the tables contain the results for the relativistic ($\sim \alpha^4$) recoil corrections given by
Eqs.~(\ref{br0}), (\ref{br1}), and (\ref{br2})-(\ref{br2end}), respectively. The eighth line of the
tables contain the QED ($\sim \alpha^5$) recoil corrections given by Eqs.~(\ref{E51})-(\ref{E512}).
The line labeled as $m\alpha^6m/M$ contains the hydrogenic approximation of the higher-order QED
effects as given by Eq.~(\ref{E61}). The uncertainty ascribed to this correction represents our
estimation of the uncalculated QED effects of order $\alpha^6$ and higher. Finally, the lines
labelled as ``Hfsmix" and ``NPol" contain the hyperfine mixing correction defined by
Eq.~(\ref{hfsmix}) and the nuclear polarizability correction given by Eq.~(\ref{np}), respectively.

The uncertainty of the total theoretical values of the isotope shift for the point-like nucleus
comes from the uncalculated higher-order QED effects. Note that our estimation of this uncertainty
is much larger than the one reported previously in Ref.~\cite{morton:06:cjp}.

%%%%%%%%%%%%%%%%%%%%%%%%%%%%%%%%%%%%%%%%%%%%%%%%%%%%%%%%%%%%%%%%%%%%%%%
\begin{table}
\caption{
$^3$He--$^4$He isotope shift of the $2^1S-2^3S$ transition,
for the point-like nucleus, in kHz.
\label{tab:is}
}
\begin{ruledtabular}
  \begin{tabular}{l.}
$\mu\,\alpha^2\,$         &       -8\,632\,567x.86 \\
$\mu\,\alpha^2\,(\mu/M)$  &           608\,175x.58 \\
$\mu\,\alpha^2\,(\mu/M)^2$&            -7\,319x.80 \\
$\mu\,\alpha^2\,(\mu/M)^3$&                  0x.30 \\
$\mu\,\alpha^4\,$         &            -8\,954x.22 \\
$\mu\,\alpha^4\,(\mu/M)$  &             6\,458x.23 \\
$\mu\,\alpha^4\,(\mu/M)^2$&                  1x.84 \\
$m\,\alpha^5\,(m/M)$      &                 56x.61   \\
$m\,\alpha^6\,(m/M)$      &                  2x.75\,(69)  \\
Hfsmix                    &                 80x.72       \\
NPol                        &                  0x.20\,(2)  \\
Sum                       &       -8\,034\,065x.66\,(69) \\
Ref.~\cite{drake:10}$^a$  &       -8\,034\,067x.8\,(1.1)\\
  \end{tabular}
\end{ruledtabular}
$^a\ $ Corrected by adding the triplet-singlet hfs mixing contribution, in order to compensate the difference in the definitions of the isotope shift.
\end{table}
%%%%%%%%%%%%%%%%%%%%%%%%%%%%%%%%%%%%%%%%%%%%%%%%%%%%%%%%%%%%%%%%%%%%%%%

%%%%%%%%%%%%%%%%%%%%%%%%%%%%%%%%%%%%%%%%%%%%%%%%%%%%%%%%%%%%%%%%%%%%%%%
\begin{table}
\caption{
$^3$He--$^4$He isotope shift of the $2^3S-2^3P$ transition,
for the point-like nucleus, in kHz.
\label{tab:is:2}
}
\begin{ruledtabular}
  \begin{tabular}{l.}
$\mu\,\alpha^2\,$         &                 12\,412\,458x.1 \\
$\mu\,\alpha^2\,(\mu/M)$  &                 21\,243\,041x.3 \\
$\mu\,\alpha^2\,(\mu/M)^2$&                    13\,874x.6 \\
$\mu\,\alpha^2\,(\mu/M)^3$&                        4x.6 \\
$\mu\,\alpha^4\,$         &                    17\,872x.8 \\
$\mu\,\alpha^4\,(\mu/M)$  &                   -20\,082x.4 \\
$\mu\,\alpha^4\,(\mu/M)^2$&                       -3x.0 \\
$m\,\alpha^5\,(m/M)$      &                      -60x.7 \\
$m\,\alpha^6\,(m/M)$      &                      -15x.5\,(3.9) \\
Hfsmix                    &                       54x.6 \\
NPol                        &                       -1x.1\,(1) \\
 Sum                      &             33\,667\,143x.2\,(3.9) \\
Ref.~\cite{morton:06:cjp}$^a$&          33\,667\,146x.2\,(7)
  \end{tabular}
\end{ruledtabular}
$^a\ $ Corrected by adding the triplet-singlet hfs mixing contribution, in order to compensate the difference in the definitions of the isotope shift.
\end{table}
%%%%%%%%%%%%%%%%%%%%%%%%%%%%%%%%%%%%%%%%%%%%%%%%%%%%%%%%%%%%%%%%%%%%%%%

The difference of the theoretical point-nucleus values and the experimental isotope shift results
should come solely from the finite nuclear size effect. The finite nuclear size correction to the
centroid energy can be parameterized as
\begin{align}
E_{\rm fs} = C\,R^2\,,
\end{align}
with $C$ being a coefficient to be calculated numerically. According Eq.~(\ref{fns}), the
coefficient $C$ has some dependence on $R$, but it is very weak and can be neglected at the level
of present interest.

The results of our calculations of the coefficient $C$ for the $^3$He-$^4$He isotope shift of the
$2^3P$-$2^3S$ and $2^1S$-$2^3S$ transition (centroid) energies are
\begin{eqnarray} \label{C1}
C(2^3P-2^3S) &=& -1212.2\,(1)\,\,\, {\rm kHz/fm}^2\,,\\
C(2^1S-2^3S) &=& -214.66\,(2)\,\,\, {\rm kHz/fm}^2\,.
\label{C2}
\end{eqnarray}
These results can be compared with the previously reported values $C(2^3P-2^3S) =
-1209.8$~\cite{morton:06:cjp} and $C(2^1S-2^3S) = -214.40$~\cite{drake:05}.

\section{Nuclear charge radii difference}

We now turn to the determination of the nuclear charge radii difference from the isotope shift.
Specifically, by comparing theoretical and experimental values of the centroid energies of
different transitions in $^3$He and $^4$He, we extract the difference of the squares of the nuclear
charge radii,
\begin{align}
\delta R^2 = R^2(^3\mbox{\rm He}) - R^2(^4\mbox{\rm He})\,.
\end{align}

Using the experimental results for the $2^1S-2^3S$ transition energies in $^3$He and $^4$He from
Ref.~\cite{rooij:11} and taking into account the experimental hyperfine shift of the $ 2^3S_1$
state, we obtain $\delta R^2$ as described in Table~\ref{tab:rms:roij}. The results is
\begin{align} \label{r2rooij}
\delta R^2[\mbox{\rm Rooij 2011}] = 1.028\,(11)\ \mbox{\rm fm$^2$}\,.
\end{align}
This value is somewhat different from the original result of Ref.~\cite{rooij:11}, $ \delta R^2 =
1.019\,(11)~\mbox{\rm fm}^2$ because of the updated theoretical prediction for the $2^1S-2^3S$
isotope shift (see Table~\ref{tab:is}).

The second experiment on the helium isotope shift suitable for the determination of $\delta R^2$ is
the measurement of the $2^3P$-$2^3S$ transition energies in $^3$He and $^4$He in
Refs.~\cite{cancio:04,cancio:12:3he}. Using these experimental values and the theory summarized in
Table~\ref{tab:is:2}, we obtain (see Table~\ref{tab:rms:cancio} for details)
\begin{align}\label{r2cancio}
\delta R^2[\mbox{\rm Cancio 2012}] = 1.074\,(4)\ \mbox{\rm fm$^2$}\,,
\end{align}
which deviates from the result (\ref{r2rooij}) by about 4$\sigma$.

The third experiment that can be used for the determination of $\delta R^2$ with comparable
accuracy is the measurement of the isotope shift difference $E(^3{\rm He},2^3P_0^{1/2} -
2^3S_1^{3/2}) - E(^4{\rm He},2^3P_2 - 2^3S_1)$ performed Ref.~\cite{shiner:95}. In this experiment,
centroid transition energies were not measured, so further theoretical and experimental input is
required to extract the charge radii difference $\delta R^2$. In order to deduce the experimental
value for the isotope shift of the $2^3P$-$2^3S$ centroid  energy, we subtract from the
experimental result of Ref.~\cite{shiner:95} the experimental hyperfine shift $\delta E_{\rm
hfs}(2^3S_1^{3/2})$, theoretical fine shift $\delta E_{\rm fs}(2^3P_0)$, and the theoretical fine
and hyperfine shift $\delta E_{\rm hfs}(2^3P_0^{1/2})$. Having obtained the isotope shift of the
$2^3P$-$2^3S$ centroid  energy, we deduce the charge radii difference $\delta R^2$ with the help of
the theory in Table~\ref{tab:is:2}. The whole determination is summarized in Table~\ref{tab:rms}.
The result is
\begin{align} \label{r2shiner}
\delta R^2 [\mbox{\rm Shiner 95}] = 1.066\,(4)\; {\rm fm}^2\,.
\end{align}

We observe that the two results from the $2^3P-2^3S$ transitions, (\ref{r2cancio}) and
(\ref{r2shiner}), are in reasonable (although, marginal) agreement with each other, but deviate
significantly from the result deduced from the $2^1S-2^3S$ transition (\ref{r2rooij}). The reason
of this discrepancy is not known at present.

%%%%%%%%%%%%%%%%%%%%%%%%%%%%%%%%%%%%%%%%%%%%%%%%%%%%%%%%%%%%%%%%%%%%%%%

\begin{table*}
\caption{Determination of the charge radii difference $\delta R^2$ from the measurement by Rooij {\em et al.} in Ref.~\cite{rooij:11},
in kHz. The remainder $\delta E$ is proportional to $\delta R^2$, $\delta E = C\, \delta R^2$,
with the coefficient $C$ given by Eq.~(\ref{C1}), see text.
\label{tab:rms:roij}
}
\begin{ruledtabular}
  \begin{tabular}{l.l}
$E(^3{\rm He},2^1S^{F=1/2} - 2^3S^{F=3/2}) - E(^4{\rm He},2^1S - 2^3S)$ &  -5\,787\,719x.2(2.4) &Exp. \cite{rooij:11}\\
$\delta E_{\rm hfs}(2^3S^{3/2})$& -2\,246\,567x.059(5) & Exp. \cite{schluesser:69,rosner:70}\\
$-\delta E_{\rm iso}(2^1S - 2^3S)$ (point nucleus) &8\,034\,065x.66\,(69) & Theory, Table~\ref{tab:is} \\[1ex]
$\delta E$                     &-220x.6(2.4) &
  \end{tabular}
\end{ruledtabular}
\end{table*}

%%%%%%%%%%%%%%%%%%%%%%%%%%%%%%%%%%%%%%%%%%%%%%%%%%%%%%%%%%%%%%%%%%%%%%%

\begin{table*}
\caption{Determination of the charge radii difference $\delta R^2$ from the measurements
by Cancio Pastor {\em et al.} in Refs.~\cite{cancio:04,cancio:12:3he},
in kHz. The remainder $\delta E$ is proportional to $\delta R^2$, $\delta E = C\, \delta R^2$,
with the coefficient $C$ given by Eq.~(\ref{C2}), see text.
\label{tab:rms:cancio}
}
\begin{ruledtabular}
  \begin{tabular}{l.l}
$E(^3{\rm He},2^3P-2S)$ (centroid) & 276\,702\,827x\,204.8\,(2.4)   &Exp. \cite{cancio:12:3he}\\
$-E(^4{\rm He},2^3P-2S)$ (centroid) & -276\,736\,495x\,649.5\,(2.1)   &Exp. \cite{cancio:04,smiciklas:10}$^a$\\
$-\delta E_{\rm iso}(2^3P - 2^3S)$ (point nucleus) &33\,667x\,143.2\,(3.9) & Theory, Table~\ref{tab:is:2} \\[1ex]
$\delta E$                     &-1x\,301.5\,(5.0) &
  \end{tabular}
\end{ruledtabular}
$^a$ obtained by combining the $2^3S_1-2^3P_{0,1}$ transition energies from Ref.~\cite{cancio:04} and
the $2^3P_0-2^3P_2$ energy from Ref.~\cite{smiciklas:10}.
\end{table*}

%%%%%%%%%%%%%%%%%%%%%%%%%%%%%%%%%%%%%%%%%%%%%%%%%%%%%%%%%%%%%%%%%%%%%%%

\begin{table*}
\caption{Determination of the charge radii difference $\delta R^2$ from the measurement by Shiner {\em et al.} in \cite{shiner:95},
in kHz. The remainder $\delta E$ is proportional to $\delta R^2$, $\delta E = C\, \delta R^2$,
with the coefficient $C$ given by Eq.~(\ref{C2}), see text.
\label{tab:rms}
}
\begin{ruledtabular}
  \begin{tabular}{l.l}
$E(^3{\rm He},2^3P_0^{1/2} - 2^3S_1^{3/2}) - E(^4{\rm He},2^3P_2 - 2^3S_1)$ &  810x\,599.0(3.0) &Exp. \cite{shiner:95}\\
$\delta E_{\rm hfs}(2^3S_1^{3/2})$&-2\,246x\,567.059(5) & Exp. \cite{schluesser:69,rosner:70}\\
$\delta E_{\rm fs}(2^3P_2)$ &-4\,309x\,074.2(1.7) & Theory \cite{pachucki:10:hefs}\\
$-\delta E_{\rm fs,hfs}(2^3P_0^{1/2})$ &-27\,923x\,393.7\,(1.7) & Theory \cite{pachucki:12:hehfs}\\
$-\delta E_{\rm iso}(2^3P - 2^3S)$ (point nucleus) &33\,667x\,143.2(3.9) & Theory, Table~\ref{tab:is:2} \\[1ex]
$\delta E$                     &-1x\,292.8(5.2) &
  \end{tabular}
\end{ruledtabular}
\end{table*}

\section{Conclusion}

In the present work we have summarized the theory of the isotope shift of the centroid energies of
light few-electron atoms. Numerical results are reported for the isotope shift of the $2^3P$-$2^3S$
and $2^1S$-$2^3S$ transition energies in $^3$He and $^4$He. As compared to the previous evaluations
\cite{morton:06:cjp,drake:10}, the higher-order recoil [$\mu\,\alpha^2\,(\mu/M)^3$ and
$\mu\,\alpha^4\,(\mu/M)^2$] and the nuclear polarizability corrections were accounted for and the
higher-order QED effects [$m\,\alpha^6\,(m/M)$] were estimated more carefully. By comparing the
theoretical predictions with the experimental results for the isotope shift, we have determined the
the difference of the squares of the nuclear charge radii of $^3$He and $^4$He, $\delta R^2$. The
results for $\delta R^2$ derived from the $2^3P$-$2^3S$ and $2^1S$-$2^3S$ transitions are shown to
disagree with each other by about 4 standard deviations. The reason of this disagreement is
presently not known.

%%%%%%%%%%%%%%%%%%%%%%%%%%%%%%%%%%%%%%%%%%%%%%%%%%%%%%%%%%%%%%%%%%%%%%%%%%%%%%%%%%%%%%
\section*{Acknowledgement}
Authors would like to acknowledge support by NCN grant 2012/04/A/ST2/00105. V.A.Y. was also
supported by the Russian Federation program for organizing and carrying out the scientific
investigations.

%\bibliographystyle{../bibtex/phaip30}
%\bibliography{../bibtex/hfst}

\end{document}